# Valley-dependent giant orbital moments and transport feature in rhombohedral graphene multilayers

Xingchi Mu, Jian Zhou*

*Center for Alloy Innovation of Design, State Key Laboratory of Mechanical Behavior of Materials, Xi'an Jiaotong University, Xi'an 710049, China*

*Email: jianzhou@xjtu.edu.cn

Abstract

Recent years have witnessed a great interest in orbital related electronics (also termed as orbitronics). In the current work, we present a first-principles density functional theory calculation on the orbital magnetic moments, intrinsic orbital Hall effect, and ordinary magnetoconductivity effects in rhombohedral graphene multilayers. Our calculations suggest a giant orbital moment that arises from inter-atomic cycloid motion, reaching over 30 $\mu_B$ under an intermediate gate voltage. This leads to a valley polarization under an external magnetic field, as observed in recent experiments [Nature **623**, 41–47 (2023)]. In addition, the orbital-related transport feature exhibit significant responses that are potentially observed in experiments. We also suggest that under a periodic field driven (such as high frequency light field), the ungated graphene multilayers could host strong quantum anomalous and orbital Hall effects, engineered by the layer number. As the graphene multilayers are intrinsically nonmagnetic with negligible spin-orbit coupling, the orbital moments would not be entangled by spin-related signals. Thus, they serve as an ideal platform to conduct orbitronic measurements and utilization for next generation information read/write nanodevices.

# I. Introduction

The use of information units in condensed matter materials has been largely-developed over the past few decades, from electronics with its carrier charge character to spintronics with the spin degree of freedom of an electron [1-4]. Recently, two new promising information carriers, valley and orbital moments (dubbed valleytronics [5-11] and orbitronics [12,13], respectively), have been proposed and attracted various attention. The valleytronics that uses the electronic behaviors at the band edge (top valence or bottom conduction bands) when they reside at different valleys such as the $K$ and $K'$ points in a honeycomb lattice. These two valleys in the momentum space are well-separated and can be individually manipulated when the intervalley interactions are weak. The orbitronics, on the other hand, is reminiscent as the spintronics with their magnetic moments arise from the cycloid motion of electrons. This has been largely overlooked as the equilibrium orbital moments are usually quenched under strong and symmetric crystal field. However, recently it has been theoretically predicted and experimentally observed that the orbital-related magnetic moments can be significant in a non-equilibrium state, such as large orbital Edelstein effect [14-17], orbital Hall effect [18-24], and orbital photocurrent generation [25-27]. The entanglement and manipulations of these quantum flavors of Bloch electrons lie at the heart of future quantum information engineering. Despite of these discoveries, even though in nonmagnetic materials without a net spin polarization, the spin-orbit coupling (SOC) between orbital magnetic moments (OMM) and spin magnetic moments (SMM) would generate observable spin-related signals, such as spin Rashba-Edelstein effect [28,29], spin Hall effect [30-32], etc. These hinder the direct observation and measurement of OMM feature. In this regard, it would be intriguing to investigate materials composed by only weak elements (such as in the first two rows in the periodic table) with strong OMM responses.

In the current work, we present first-principles calculations based on density functional theory (DFT) to investigate a simple material platform with giant orbital moments, namely, graphene multilayers in rhombohedral stacking order [33-37]. Very recently, this material family has been receiving various experimental attention as it shows many exotic physical behaviors, such as multiferroic orders under gate voltage [38], tunable quantum anomalous Hall (QAH) conductance [39], and fractionalized Chern insulating feature [40]. The band dispersion at the Fermi energy of rhombohedral multilayer (except single layer) graphene usually exhibits flat band character, giving large density of states and valley-polarized Berry phase structure. Hence, the rhombohedral stacking could exhibit strong correlation effects without precise Morié pattern control. Despite these advanced discoveries, until now, first-principles DFT based quantitative evaluations of its orbitronic behaviors are still lacking, such as quantitative OMM distribution on the electronic band and the manipulation of valley polarization under external fields. Furthermore, orbital-based transport feature, such as intrinsic orbital Hall effect and ordinary magnetoconductivity remain to be disclosed. Here, we compute the band-resolved orbital moment with different layer-numbers in rhombohedral graphene multilayers under gate voltage, and investigate the orbital-related transport properties [Fig. 1(a)]. Intrinsically, the systems exhibit gapless semimetallic feature, without a net OMM protected by the simultaneous spatial inversion ($\mathcal{P}$) and time-reversal ($\mathcal{T}$) symmetry. Our calculations show that a gate field can open the bandgap and induce an orbital moment reversal in the $k$-space (near $K$ or $K'$ valleys but with opposite signs), consistent with model calculations. The magnitude of orbital moment could be enhanced as large as 30 $\mu_B$ under an intermediate field, which exceedingly surpass the magnetic moments that a single spin could carry. This arises from the inter-atomic cycloid motion in real space and is ascribed by a large interband dipole transition in $k$-space. Consequently, our calculation reveals observable band valley polarization, giant intrinsic orbital Hall effect, and ordinary magnetoconductivity in the system, owing to significant OMM values. We also suggest that a periodic light field could also open the material bandgap and induce large quantum anomalous and orbital Hall conductance.

## II. Methods

Our DFT calculations are performed within the Vienna *ab initio* simulation package (VASP) [41,42] using the solid state Perdew-Burke-Ernzerhof (PBEsol) type generalized gradient approximation (GGA) method [43] to treat the exchange-correlation interactions. Projector augmented-wave (PAW) [44] and a plane wave basis set with kinetic cutoff energy of 500 eV are adopted to describe the core and valence electrons, respectively. In order to perform integrals in the first Brillouin zone (BZ), we use Γ-centered special $k$-mesh with (45×45×1) grids [45]. These parameters have been carefully tested to yield converged band dispersion and orbital magnetic moments ($m^L$). As the carbon element is light weighted, we omit the SOC throughout our calculations, which was predicted to open a marginal gap ($10^{-6}$ eV) for a single layer graphene [46]. Hence, all the bands are doubly degenerated in this work. Geometric relaxations have been performed using the conjugate gradient algorithm scheme, with the van der Waals interactions treated using the Grimme's DFT-D3 with zero-damping function method [47]. The vacuum space along the z direction of over 15 Å is added to eliminate the interactions between periodic images. The gate field ($E_z$) is applied using the sawtooth electric potential scheme, with corrected long-range dipole-dipole interactions. We fit the Hamiltonian using Wannier functions composed of C-s and p orbitals, as implemented in the Wannier90 code [48]. We find that the intercell interactions in multilayer graphene systems are very large (over ~30 unit cells), indicating that the cycloid motion of Bloch electrons could yield large orbital moments.

### III. Results and Discussions.

#### A. Geometric and electronic structures of multilayer graphene

We take a pentalayer rhombohedral graphene thin film to illustrate our calculation results in the main text, and the variation under different layer numbers (for bilayer, trilayer, and tetralayer) are mainly reported in Supplemental Material (SM) [49]. Figure 1(b) presents its relaxed geometric structure, with ultraflat shape of each layer. The interlayer distance is optimized to be 3.44 Å. The relaxed system belongs to $p\bar{3}m1$ layer group with an inversion center at the middle of the third layer. The intact band dispersion along the high symmetric **k**-path is plotted in Fig. 1(c), with its $p_z$ orbital resolved contribution color-coded. One sees that there are ten orbitals near the Fermi level, consistent with its pentalayer structural nature. In the vicinity of *K* and *K'* valleys, only $p_z$ orbital character appears, consistent with the single layer graphene π orbital character. When we zoom into both valleys within –15 ~ 15 meV energy window [Fig. 1(d)], only a valence and a conduction band appear, with tilted parabolic dispersion relation. This is disparate from the bilayer graphene system (Fig. S1) [49]. Under the simultaneous $\mathcal{P}$ and $\mathcal{T}$ symmetry, the orbital moments of each state are exactly zero, as $\mathcal{P}\boldsymbol{m}^L(k_x,k_y) = \boldsymbol{m}^L(-k_x,-k_y)$ and $\mathcal{T}\boldsymbol{m}^L(k_x,k_y) = -\boldsymbol{m}^L(-k_x,-k_y)$. This is also verified by our numerical calculations (not shown).

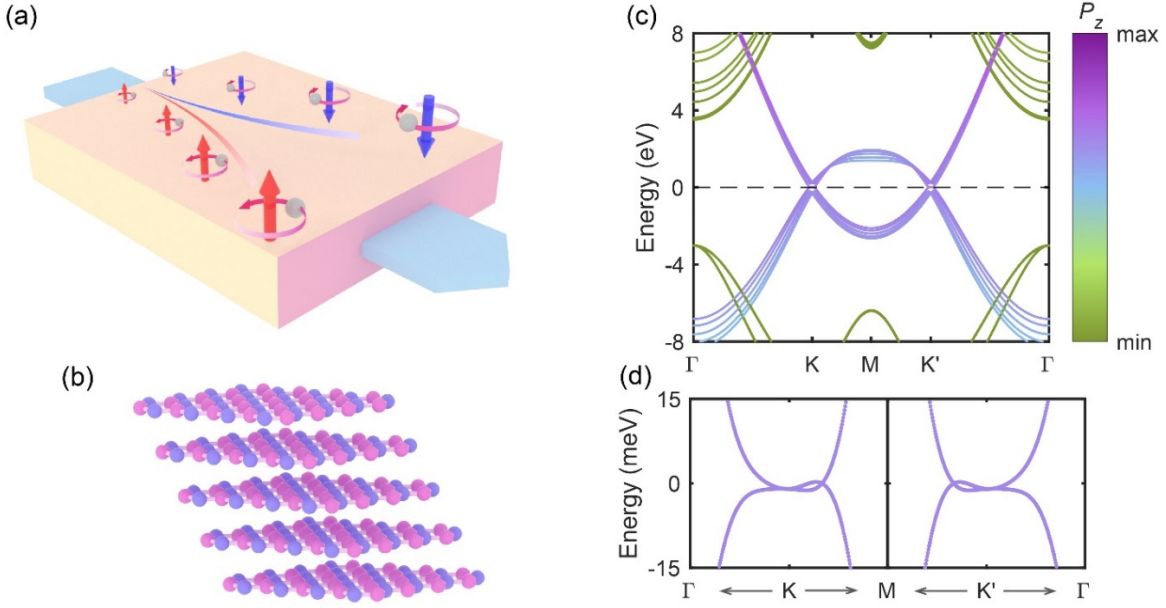

FIG. 1. (a) The schematic diagram of the orbital Hall effect with the red and blue arrows representing positive and negative OMM states, respectively. (b) Atomic geometric and (c) electronic band dispersion of a pentalayer rhombohedral graphene thin film. The colormap represents the $p_z$ orbital contribution portion of each state. (d) shows enlarged band dispersions near the $K$ and $K'$ valleys, respectively.

### B. Gate voltage tuned band structure and orbital moments

In order to induce finite orbital moment, one could break either $\mathcal{P}$ or $\mathcal{T}$ in the system. Now we apply a gate electric field ($E_z$) that has been adopted in experiments [38]. Figure 2 shows a typical band dispersion and the $z$-component orbital moment for each state. One clearly sees that the bandgap opens at the $K$ ($K'$) valley under $E_z$, and both top valence and bottom conduction bands (TVB and BCB) exhibit a clear Mexican-hat shape. According to the modern theory [50-53], the topological trivial OMM can be evaluated according to

$$m_{n\mathbf{k}}^L = \frac{e}{2\hbar}\mathfrak{I}[\langle \nabla_{\mathbf{k}} u_{n\mathbf{k}}|\times (H_{\mathbf{k}} - \varepsilon_{n\mathbf{k}})|\nabla_{\mathbf{k}} u_{n\mathbf{k}}\rangle] \quad (1)$$

Here, $|u_{n\mathbf{k}}\rangle$ and $\varepsilon_{n\mathbf{k}}$ are the eigenfunction and eigenenergy of Bloch state at band $n$ and momentum $\mathbf{k}$. $H_{\mathbf{k}}$ is Hamiltonian at $\mathbf{k}$. Owing to the ultrathin two-dimensional (2D) nature, the in-plane orbital moment components are ill-defined, which will not be discussed in this work. Hence, we only focus on its $z$ component $m^{L_z}$. Intuitively, the $p_z$ ($= Y_1^0$ in spherical harmonic function expression) atomic orbital would not exhibit nonzero $m^{L_z}$, as its magnetic orbital quantum number is zero. We also perform numerical simulation by fitting the DFT bands with *s* and *p* atomic wavefunctions, and compute the intra-atomic contributed OMM, giving almost vanishing values. However, according to the modern theory [Eq. (1)] that incorporates inter-atomic orbital cycloid motion, our calculation shows that each Bloch state exhibits magnetic moments as large as a few tens of Bohr magneton ($\mu_B$). For instance, when the $E_z$ is 0.08 V/Å [Fig. 2(a), corresponding to 1.37 V in gate voltage], the orbital moments of TVB at the *K* point are positive (5.0 $\mu_B$), and they become negative (–5.0 $\mu_B$) at *K'*, protected by $\mathcal{T}$. Interestingly, in the vicinity of both valleys, the orbital moments reverse their signs at the Mexican-hat band edge. On the contrary to low energy models, here the band dispersion lacks of particle-hole symmetry, due to the correlation effect and long range orbital interactions in the *ab initio* calculations. The largest OMM reaches 23.7 $\mu_B$ according to our calculation. This is significantly large, especially compared with SMM contribution which is limited to be –1 ~ 1 $\mu_B$. If the gate field direction is flipped [Fig. 2(b)], the band energy remains the same, but the orbital moments near each valley exchange their signs. This can be understood as the $E_z$ reversal corresponds to a spatial inversion mapping between the two states.

We would like to note that such orbital moment sign variation along the Mexican-hat is not a ubiquitous feature. If $E_z$ is furthermore enhanced, such as doubled to 0.16 V/Å, we find that the sign of $m^{L_z}$ in the vicinity of each valley keeps to be the same [Fig. 2(c)]. This actually arises from the orbital inversion between the TVB and TVB–1 (one band below), and between the BCB and BCB+1 (one band above). The largest (smallest) OMM values become 30.2 (–30.2) $\mu_B$, indicating a large electromagnetic coupling in graphene multilayer systems. We plot the ***k***-path dependent $m^{L_z}$ for the TVB near the *K* valley in Fig. 2(d). For the other valley (*K'*), their signs will flip under $\mathcal{T}$. We plot the results of bilayer to tetralayer graphene systems in Figs. S2 and S3 [49], which exhibit similar behavior as in the pentalayer graphene. The maximum OMM in these cases could all reach ~30 $\mu_B$.

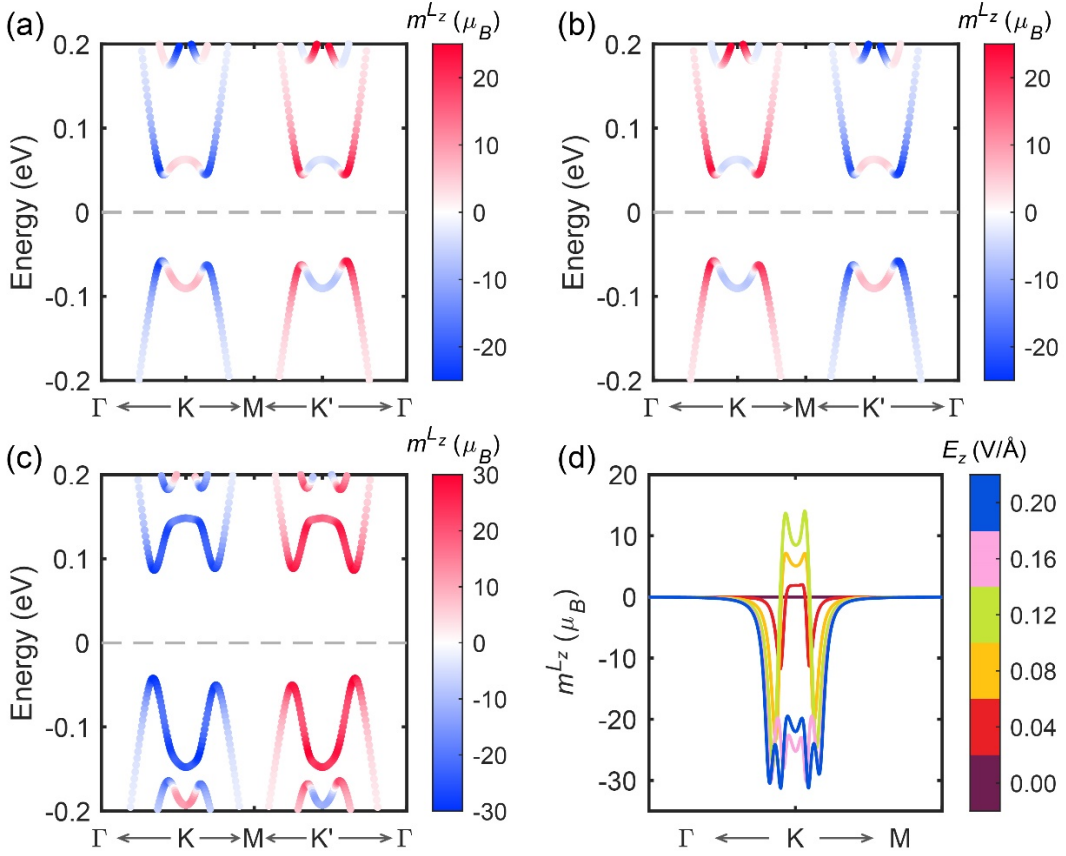

FIG. 2. Vertical electric field $E_z$ induced bandgap opening and orbital moments. (a) – (c) plot the band dispersion along the high symmetric $\boldsymbol{k}$-path with $E_z$ = 0.08 V/Å, –0.08 V/Å, and 0.16 V/Å, respectively. (d) shows $m^{L_z}$ variation for the TVB in the vicinity of $K$ valley as $E_z$ increases from 0 to 0.20 V/Å. The negative $E_z$ situations are not plotted.

### C. Intrinsic orbital Hall conductance

The transport of OMM, such as orbital Hall effect (OHE) and orbital photocurrent generations, lies at the heart of orbitronics technology. As the state-resolved OMM is significant under $E_z$, we further carry out calculations to evaluate the intrinsic anomalous Hall effect and orbital Hall effect (OHE). We calculate the Berry curvature and orbital Berry curvature distribution in the momentum space. According to the Kubo perturbation theory, the Berry curvature is

$$\Omega_z(\boldsymbol{k}) = \sum_{n \neq m}(f_{n\boldsymbol{k}} - f_{m\boldsymbol{k}})\frac{\hbar^2 \Im \langle u_{m\boldsymbol{k}}|v_x|u_{n\boldsymbol{k}}\rangle\langle u_{n\boldsymbol{k}}|v_y|u_{m\boldsymbol{k}}\rangle}{(\varepsilon_{n\boldsymbol{k}} - \varepsilon_{m\boldsymbol{k}})^2}. \qquad (2)$$

Here, $f_{n\boldsymbol{k}}$ is the Fermi-Dirac distribution function, and $v_x$ ($v_y$) refers to velocity operator. From Fig. 3(a), the Berry curvature mainly localizes at the two valleys, with opposite signs. This indicates a valley-dependent anomalous Hall conductance in the gated graphene multilayers. The total anomalous Hall effect is absolutely zero, protected by $\mathcal{T}$. If we introduce OMM degree of freedom into consideration, one can evaluate intrinsic OHE by integrating orbital Berry curvature, through replacing the operator $v_y$ in Eq. (2) by an anticommutation orbital current operator $\{v_y, L_z\} = v_y L_z + L_z v_y$, $L_z$ is orbital angular momentum, which is described as $L_z = -(\hbar/\mu_B g_L) m^L$. Landé $g$-factor $g_L = 1$ is taken because the SOC is neglected due to low carbon mass. Note that if the graphene is deposited onto substrate with heavier elements, then its effective Landé $g$-factor will be enhanced under proximity effect. The off-diagonal elements is evaluated according to $\langle n|L_z|m\rangle = \frac{e\hbar}{2g_L\mu_B}\sum_{l\neq m,n}\{\langle n|v_x|l\rangle\langle l|\mathcal{A}_y|m\rangle + [\langle n|v_x|n\rangle + \langle m|v_x|m\rangle]\langle n|\mathcal{A}_y|m\rangle - \langle x \leftrightarrow y\rangle\}$, where $\langle n|\mathcal{A}_y|m\rangle = i\langle n\boldsymbol{k}|\partial_{k_y}|m\boldsymbol{k}\rangle$ is the interband Berry connection. Hence, the intrinsic OHE conductance is evaluated via

$$\sigma_x^{y,L_z} = e\int d^2\boldsymbol{k}\,\Omega_z^{L_z}(\boldsymbol{k}) = e\int d^2\boldsymbol{k}\sum_{n\neq m}(f_{n\boldsymbol{k}} - f_{m\boldsymbol{k}})\frac{\hbar^2\Im\langle u_{m\boldsymbol{k}}|v_x|u_{n\boldsymbol{k}}\rangle\langle u_{n\boldsymbol{k}}|\{v_y,L_z\}|u_{m\boldsymbol{k}}\rangle}{(\varepsilon_{n\boldsymbol{k}}-\varepsilon_{m\boldsymbol{k}})^2}. \quad (3)$$

Note that recently there have been some discussions on the intrinsic and extrinsic contributions in orbital Hall conductance for doped systems [54,55]. In the current work, as the graphene multilayer is composed by only carbon atoms and can be synthesized in a very good sample quality, the intrinsic part of Hall conductance dominates over the disorder or impurity induced extrinsic part. Hence, we only focus on the intrinsic OHE here, while leaving the extrinsic contributions (such as skew effect and side jump) and orbital torque effect [56,57] (arising from the non-conservation feature of orbital quantum number) in the future. As can be seen in Fig. 3(b), the two valleys contribute $\Omega_z^{L_z}(\boldsymbol{k})$ with the same sign. This arises from the fact that $\Omega_z^{L_z}(\boldsymbol{k})$ is invariant under both $\mathcal{P}$ and $\mathcal{T}$. We also see some small contributions near the $\Gamma$ point. Under $\mathcal{C}_{3z}$ rotation, the valley-dependent $\Omega_z^{L_z}(\boldsymbol{k})$ shows slight three-fold warping texture. Figure 3(c) illustrates the chemical potential ($\mu$) dependent OHE conductance, under different $E_z$. One sees a clear horizontal plateau when the chemical potential lies within the bandgap. When $E_z$ = 0.04 V/Å, the orbital Hall conductance $\sigma_x^{y,L_z}$ inside bandgap reaches 557 $e/2\pi$. This value is larger than the second-order topological insulator Janus RuBrCl (1.30 $e/2\pi$) [58] and monolayer MoS$_2$ (2.6 $e/2\pi$) [59], according to previous works. When $E_z$ is enhanced, the bandgap increases, leading to smaller $\sigma_x^{y,L_z}$ height magnitude [Eq. (3)] but with wider plateau regime. According to our calculation, under an intermediate gate field, the smallest magnitude is 107 $e/2\pi$ with $E_z = 0.12$ V/Å, which is still much larger than the second-order topological insulator Janus RuBrCl and monolayer MoS$_2$. Note that flipping $E_z$ would not change the sign and value of $\sigma_x^{y,L_z}$, as it is a $\mathcal{P}$ invariant variable. Similar $E_z$ modulated intrinsic OHE plateau can be observed in bilayer, trilayer, and tetralayer graphene systems, as plotted in Fig. S4 [49]. We find that a small $E_z = 0.04$ V/Å would yield an OHE conductance of 294, 350, and 473 $e/2\pi$ inside the bandgap, respectively. One may notice that the OHE inside the bandgap seems increasing under thicker systems. In order to reveal the thickness effect, we take a typical voltage value (~1.1 V) and explore the variation of bandgap and OHE as functions of layer number $N$. Our calculations suggest that the bandgap remains reduced as layer number increases until $N = 7$, giving a monotonously increasing OHE conductance. However, at $N = 8$, the optimized geometry starts to show some $sp^3$ hybridization feature near the Fermi level, leading to a wider bandgap and a smaller OHE. Furthermore, our geometric calculations show that they would conduct a phase transformation towards diamond-like structure with very large $N$. Hence, there exists a saturation layer number (at a chosen gate voltage level), at which the bandgap reaches a minimum value with a large OHE inside the gap.

We also note that the experimental detection and measurement of OHE are receiving tremendous attention recently, even though still facing challenges of distinguishing it from spin signals. In some light-element materials with marginal SOC effect, OHE has been measured via magneto-optical Kerr effect and orbital torque effect, such as metal Ti [22], Cr [21] and Si [23]. Therefore, the OHE detection in rhombohedral multilayer graphene can be conducted in the similar manner.

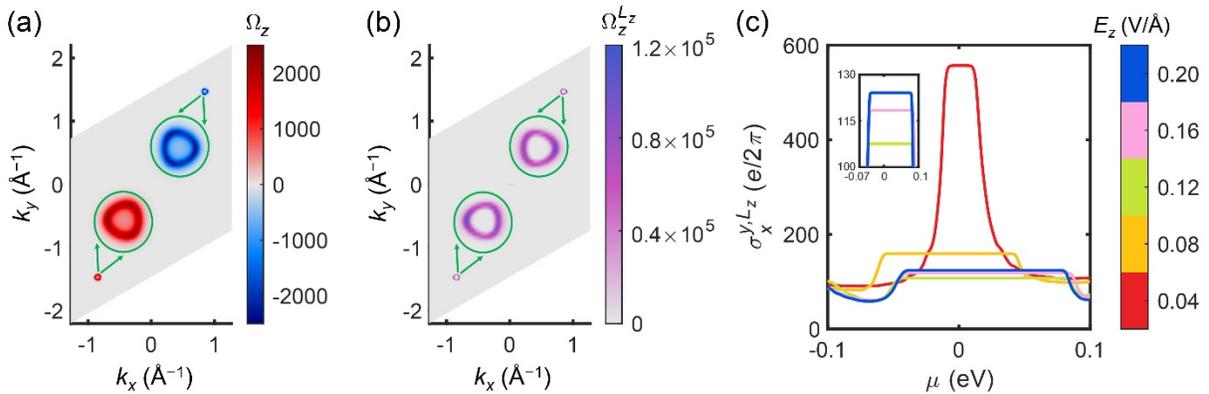

FIG. 3. (a) Berry curvature and (b) orbital Berry curvature distribution in the momentum space under $E_z = 0.08$ V/Å, with the chemical potential lies in the bandgap of pentalayer graphene. The green circular inset shows an enlarged view of the Berry curvature or orbital Berry curvature distribution at $K/K'$ valley. (c) Intrinsic orbital Hall effect under different $E_z$. The abscissa axis denotes the electronic chemical potential ($\mu$) with respect to the intrinsic Fermi energy. Inset shows enlarged plot under stronger $E_z$.

### D. Magnetic field induced valley polarization and ordinary magnetoconductivity

In order to lift the valley degeneracy, one could apply a vertical magnetic field ($B_z$) that breaks $\mathcal{T}$. We estimate the valley energy difference by the Zeeman exchange interaction [51] [depicted in Fig. 4(a)],

$$\tilde{\varepsilon}_{n\mathbf{k}}(B_z) = \varepsilon_{n\mathbf{k}} - B_z m_{n\mathbf{k}}^{L_z}. \tag{4}$$

As the band dispersion shows a Mexican-hat shape with variable $m_{n\mathbf{k}}^{L_z}$, the valley energy difference (between $K$ and $K'$ valleys) at the TVB and BCB are shown in Figs. 4(b) and 4(c), respectively. It is clear that the variation of $\tilde{\varepsilon}_{n,K} - \tilde{\varepsilon}_{n,K'}$ exhibits a linear relationship with its slope proportional to $m_{nK}^{L_z} - m_{nK'}^{L_z}$. Hence, under larger $E_z$, the valley splitting becomes more significant. Our estimation suggests that a 1 T magnetic field would yield $\tilde{\varepsilon}_{VB,K} - \tilde{\varepsilon}_{VB,K'}$ (or $\tilde{\varepsilon}_{CB,K} - \tilde{\varepsilon}_{CB,K'}$) ~3 meV. One may wonder if the opposite OMM at two valleys could generate a band inversion and yield valley-polarized QAH effect, as predicted in magnetic materials with strong inversion symmetry breaking. According to our calculations, under weak $E_z$, both the bandgap and OMM are small, and such a band inversion cannot occur under moderate $B_z$. Increasing $E_z$ could enhance both OMM and bandgap, so that the band inversion is still unlikely to happen. Hence, the total anomalous Hall effect, under both $E_z$ and $B_z$, remains to be zero with valley dependent nature.

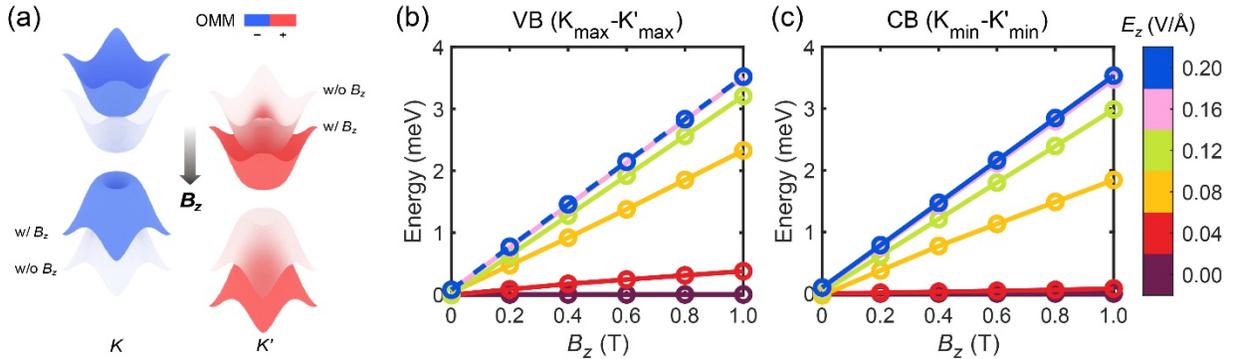

FIG. 4. (a) Schematic plot that illustrates the energy degeneracy lift of the two valleys under a magnetic field, arising from their distinct orbital magnetic moment distributions. The blue and red colors represent negative and positive orbital magnetic moments (OMM), respectively. The light and dark colors indicate without and with magnetic field $B_z$, respectively. Valley difference of (b) TVB maximum and (c) BCB minimum energy variation under Zeeman exchange effect of a vertical magnetic field $B_z$.

Then we evaluate the OMM transport feature, namely, ordinary magnetoconductivity (OMC) effect [60-62], as the Berry curvature-induced anomalous Hall conductance is always zero. When one considers the Lorentz force effect of the vertical $B_z$ field, the OMC can be estimated according to [63-67]

$$\sigma_{ij}^{\mu}(B_z) = \frac{e^2}{4\pi^3} \int d\mathbf{k} \sum_n \tau_n v_{n\mathbf{k}}^i \tilde{v}^j{}_{n\mathbf{k}} \left(-\frac{\partial f}{\partial \varepsilon}\right)_{\varepsilon=\varepsilon_{n\mathbf{k}}}, \tag{5}$$

where the integral is taking along the $\mathbf{k}$-path $\frac{d\mathbf{k}_n(t)}{dt} = -\frac{e}{\hbar} \mathbf{v}_{n\mathbf{k}} \times \mathbf{B}$, $\tilde{\mathbf{v}}_{n\mathbf{k}} = \int_{-\infty}^{0} \frac{dt}{\tau_n} e^{\frac{t}{\tau_n}} \mathbf{v}_{n\mathbf{k}}$, and $f = \frac{1}{e^{(\varepsilon-\mu)/k_B T}+1}$ ($T$ refers to temperature). Here, $\tau_n$ is the band-resolved carrier relaxation time. Note that an exact estimate of $\tau_n$ in perfect crystal is not straightforward, which arises from electron-phonon interaction and electron-electron scattering. In a realistic material, it strongly depends on the quality sample, impurity, disorder, and environmental effects. Hence, following previous works [63,65-67], we assume a universal relaxation time approximation, which is denoted as $\tau$.

We take a typical case with $E_z$ = 0.08 V/Å and evaluate OMC under different temperature (in the range of 125−325 K, see Fig. 5). Following previous works, the $\sigma \cdot \tau$ is plotted under the variation of $B_z \tau$. The chemical potential is tuned slightly above the conduction band minimum [Figs. 5(a) and 5(b)] and below the valence band maximum [Figs. 5(c) and 5(d)], respectively. Isotropic OMC can be seen, with $\sigma_{xx} = \sigma_{yy}$ under $\mathcal{C}_{3z}$ symmetry while the nondiagonal components satisfy $\sigma_{xy} = -\sigma_{yx}$. One can see that as the temperature is enhanced, the magnitude of longitudinal and Hall magnetoconductivity reduce. In both cases, there are electron pocket and hole pocket contributions, due to the Mexican-hat shape band edge. According to the well-known two-band model [65,67-69], one can approximate the magnetoconductivity under electron and hole carriers as

$$\sigma_{xx} = e \left[\frac{n_e \mu_e}{1+(\mu_e B)^2} + \frac{n_h \mu_h}{1+(\mu_h B)^2}\right] \tag{6}$$

$$\sigma_{xy} = eB \left[\frac{n_h \mu_h^2}{1+(\mu_h B)^2} - \frac{n_e \mu_e^2}{1+(\mu_e B)^2}\right], \tag{7}$$

where $n_e$ and $n_h$ denote the electron and hole carrier concentration, respectively. $\mu_e$ and $\mu_h$ are their mobility. Therefore, the longitudinal magnetoconductivity $\sigma_{xx}$ reduces monotonously with magnetic field strength. At the weak $B$-field limit, $\sigma_{xx}$ reduces as $B^2$, while in the strong $B$-field limit, $\sigma_{xx} \sim 1/B^2$. The Hall-like magnetoconductance $\sigma_{xy} \sim B$ (linear dependence) in the weak $B$-field limit, and it shows saturation to a constant value as $\sigma_{xy} \sim 1/B$ with strong $B$. If the dominant carrier is electron (hole), $\sigma_{xy}$ is negative (positive). These facts can be clearly seen in our simulation results. When the system is $n$-doped [Fig. 5(b)], the calculated $\sigma_{xy}$ is negative, suggesting electron dominated carriers. On the contrary, if the chemical potential is moving below the valence band [Fig. 5(d)], $\sigma_{xy}$ becomes positive with dominant hole contributions. Note that in previous works, Lau *et al.* measured magnetoconductance in weakly doped trilayer rhombohedral graphene [60]. The variation of conductance under magnetic field is found to be negative under higher temperature, which is consistent with our computational results.

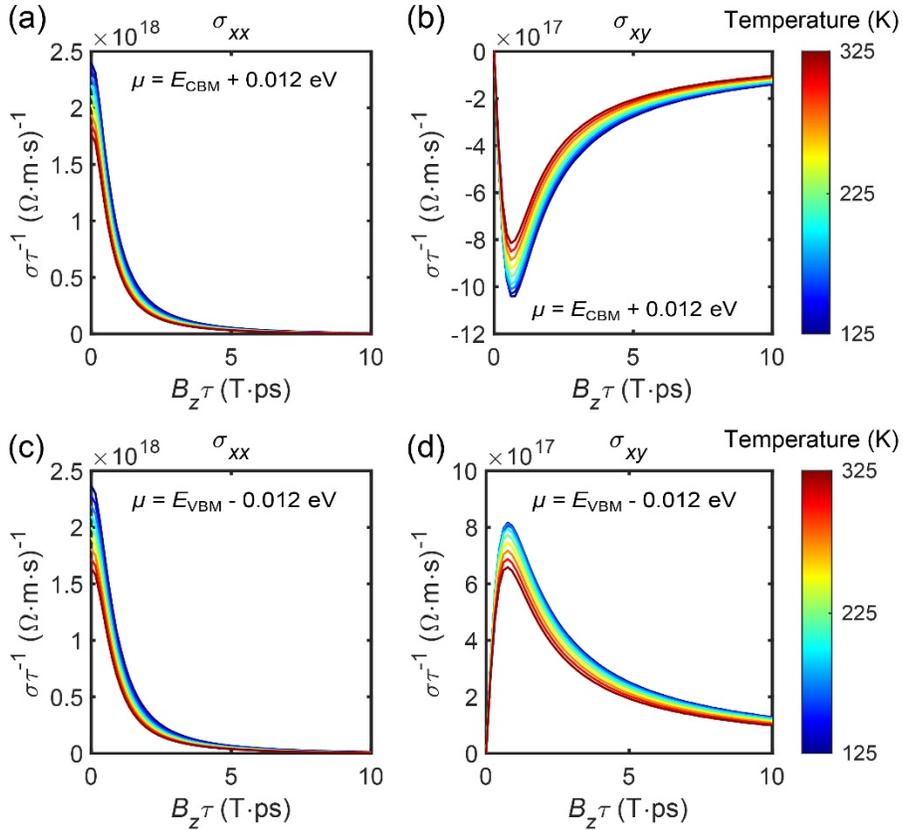

FIG. 5. (a) Longitudinal and (b) Hall conductance components of the ordinary magnetoconductivity, when the chemical potential ($\mu$) is shifted slightly above the conduction band

minimum in pentalayer rhombohedral graphene. (c) and (d) show the results with $\mu$ lies below the valence band maximum. In both cases, a gate field of $E_z = 0.08$ V/Å is applied.

### E. Light-driven quantum anomalous and orbital Hall effects

In addition to breaking $\mathcal{P}$-symmetry using a gate field, one can also break $\mathcal{T}$-symmetry to open finite bandgap and induce nonzero OMM at each band. Here, we discuss the circularly polarized light (CPL) irradiation with high frequency well-above the bandgap energy regions. Such a time-periodic field effect can be described by Floquet band engineering [70-78]. Under the Peierls substitution, the Hamiltonian is replaced to be $H(\boldsymbol{k}) \to H\left(\boldsymbol{k} + \frac{e\boldsymbol{A}(t)}{\hbar}\right)$, where $\boldsymbol{A}(t) = A_0(\cos\omega t, \eta \sin\omega t, 0)$ is the light vector potential with $\omega = 8$ eV and $\eta = \pm 1$ representing angular frequency and light chirality. We adopt van Vleck-Magnus expansion [79] to obtain the Floquet Hamiltonian

$$H_{\mathrm{F}}(\boldsymbol{k}) = H_0(\boldsymbol{k}) + \sum_{m \geq 1} \frac{[H_{-m}(\boldsymbol{k}), H_m(\boldsymbol{k})]}{m\omega} + o\left(\frac{1}{\omega}\right), \tag{6}$$

where $H_m(\boldsymbol{k}) = \frac{\omega}{2\pi} \int_0^{\frac{2\pi}{\omega}} dt e^{im\omega t} H_0\left(\boldsymbol{k} + \frac{e\boldsymbol{A}(t)}{\hbar}\right)$ is the Fourier transformation component of time-dependent Hamiltonian, and $[H_{-m}(\boldsymbol{k}), H_m(\boldsymbol{k})] = H_{-m}(\boldsymbol{k})H_m(\boldsymbol{k}) - H_m(\boldsymbol{k})H_{-m}(\boldsymbol{k})$ represents the commutator operator. In our practice, according to careful tests, we only consider the first order expansion ($m = 1$), which gives well-converged results near the Fermi level.

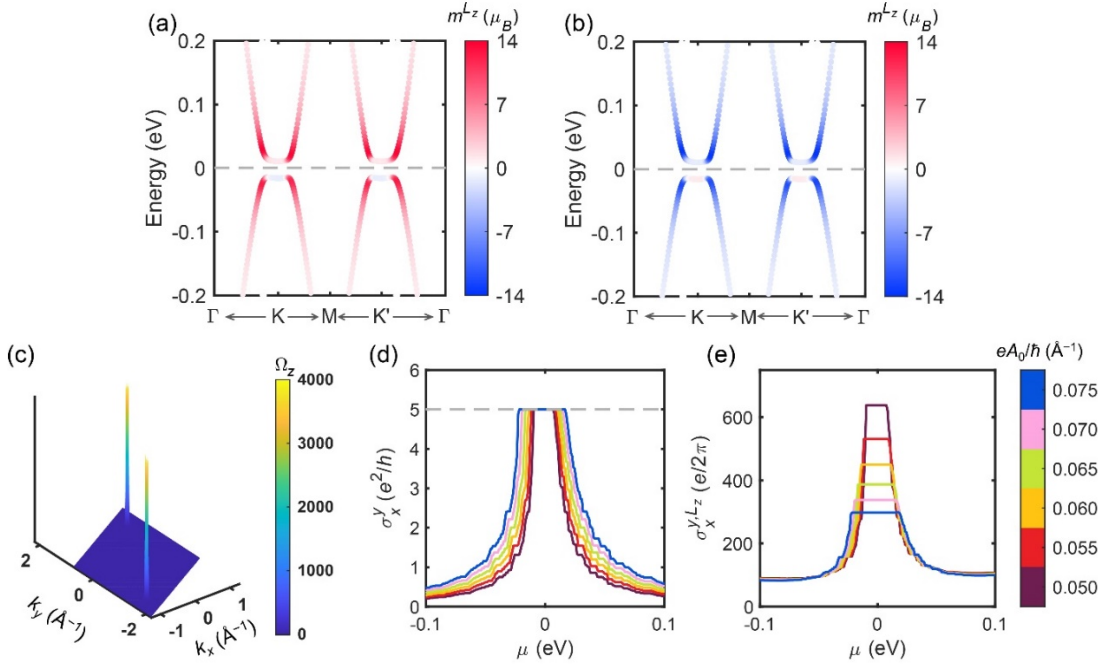

FIG. 6. The Floquet-induced bandgap opening and orbital magnetic moment for pentalayer graphene, under (a) left-handed CPL and (b) right-handed CPL with intensity of $eA_0/\hbar = 0.06$ Å$^{-1}$. (c) Berry curvature distribution in the momentum space under left-handed CPL. (d) Anomalous Hall conductance and (e) orbital Hall conductance variations under different $eA_0/\hbar$, under a left-handed light irradiation. Note that changing light handedness would flip $\sigma_x^y$ values while keep $\sigma_x^{y,L_z}$ values.

As the intact state is gapless semimetallic, a significantly weak but finite light field could open its bandgap. This has been discussed in Floquet-engineering of the topological surface states, in which one yields a QAH state with finite Chern number (its sign depends on light chirality) under an arbitrarily weak CPL irradiation. For instance, under a light intensity of $\frac{eA_0}{\hbar} = 0.06$ Å$^{-1}$ irradiating onto the pentalayer graphene, its bandgap opens with 0.025 eV [Figs. 6(a) and 6(b)]. Once again, the band dispersions show a slight Mexican-hat shape. In both cases, the band-resolved OMM arises, and they give opposite values under left- and right-handed light. The largest OMM reaches ~13 $\mu_B$, lying at the band edge at both valleys. On the contrary to the $E_z$ field where the OMM values are opposite in the two valleys, here the $K$ and $K'$ valleys exhibit the same OMM under CPL driving, protected by $\mathcal{P}$.

The bandgap opening usually indicates topological phase transition. In the current case, the existence of $\mathcal{P}$ gives $\Omega_z(\boldsymbol{k}) = \mathcal{P}\Omega_z(-\boldsymbol{k})$. We plot the Berry curvature distribution in the momentum space in Fig. 6(c), giving two sharp peaks near *K* and *K'* valleys. In addition, we calculate the anomalous Hall conductance by integrating Berry curvature [Eq. (2)], and the results are shown in Fig. 6(d). Interestingly, one sees a large Chern number of 5 under a left-handed CPL irradiation. As light intensity is enhanced, the bandgap value increases, while the QAH effect remains to be unchanged. We also note that such a large Chern number is dependent on the layer numbers [Fig. S5(a)]. Under an opposite CPL chirality, the QAH conductance would flip its sign, indicating light modulated QAH conductance. This is consistent with previous theoretical model results [77,78].

In addition to a high Chern number state, we perform calculations on the intrinsic OHE effect. Unlike QAH conductance, the OHE is not a quantized to be an integer even inside the bandgap. As depicted in Fig. 6(e), one sees that a weaker light could yield larger OHE conductance plateau, which is narrow as the bandgap is small. For instance, under $eA_0/\hbar = 0.05$ Å$^{-1}$, the in-gap OHE conductance reaches 638 $e/2\pi$, on the same order of gate field $E_z$. When the light intensity is enhanced to $eA_0/\hbar = 0.075$ Å$^{-1}$, the bandgap becomes 0.039 eV and the OHE is reduced to 298 $e/2\pi$. Note that as OHE is $\mathcal{T}$ preserved, flipping light handedness would not change its value, different from the QAH effect. We also investigate the layer number dependent OHE under Floquet bandgap engineering. As shown in Fig. S5(b), one sees that fewer layer graphene reduces the OHE conductance value, under the same CPL intensity ($eA_0/\hbar = 0.05$ Å$^{-1}$). Such an in-gap OHE conductance follows a linear relationship as the graphene layer number is varied, suggesting a thickness controlled OHE under light.

## IV. Conclusion

In this work, we perform first-principles DFT calculations to investigate the orbital moment generation and its transport feature in rhombohedral graphene multilayers. In order to induce finite bandgap in the system, the situations of a gate field breaking $\mathcal{P}$ and a CPL breaking $\mathcal{T}$ are discussed. Our calculations suggest that bandgap opening would introduce significant band-resolved OMM, reaching a few tens of $\mu_B$. The gate field would yield valley-dependent OMM distribution, so that valley polarization on the band energy can be realized using a complementary magnetic field. The OHE and OMC responses are evaluated, both giving significant values as compared with other recently studied 2D materials. Under periodic light field, QAH could arise with a large Chern number and a strong OHE conductance can be achieved. We also investigate the layer number dependent character, suggesting that thicker structure could exhibit significant OMM responses. As graphene is composed by light-weight carbon elements with marginal SOC and non-spin polarization, the rhombohedral multilayer graphene serves as an ideal platform to investigate orbitronic-related magnetic feature, without entangled by SMM signals. Considering the recent experimental advances, we expect that our predictions are potentially explored and verified experimentally.

**Acknowledgments.** This work is supported by the National Natural Science Foundation of China under Grant No. 12374065. The computations are performed in the Hefei Advanced Computing Center.

**Data Availability Statement.** The datasets generated and analyzed during the current study are available from the corresponding author upon reasonable request.